\pdfoutput=1
\documentclass[10pt,twocolumn,letterpaper]{article}

\usepackage[pagenumbers]{cvpr} % To force page numbers, e.g. for an arXiv version

\usepackage{graphicx}
\usepackage{amsmath}
\usepackage{amssymb}
\usepackage{booktabs}
\usepackage{multirow}
\usepackage{xcolor,colortbl}
\usepackage{ulem}

\usepackage[pagebackref,breaklinks,colorlinks]{hyperref}

\usepackage[capitalize]{cleveref}
\crefname{section}{Sec.}{Secs.}
\Crefname{section}{Section}{Sections}
\Crefname{table}{Table}{Tables}
\crefname{table}{Tab.}{Tabs.}

\newcommand{\tx}{\tilde{x}}
\newcommand{\tX}{\tilde{X}}
\newcommand{\hx}{\hat{x}}
\newcommand{\hM}{\hat{M}}

\newcommand{\ys}{\{y_i\}_i}
\newcommand{\ysp}{\{y_i\}_{i \in I^+}}

\newcommand{\pmr}[1]{\scriptsize$\pm$#1}

\def\cvprPaperID{5900} % *** Enter the CVPR Paper ID here
\def\confName{CVPR}
\def\confYear{2023}

\begin{document}

\title{ReVISE: Self-Supervised Speech Resynthesis with Visual Input\\for Universal and Generalized Speech Enhancement}

\author{
Wei-Ning Hsu$^{1}$, Tal Remez$^1$, Bowen Shi$^{1,3}$, Jacob Donley$^2$, Yossi Adi$^{1,4}$\vspace{.2em}\\
$^1$FAIR, Meta AI Research  \hspace{.5em} $^2$Meta Reality Labs Research \\
\hspace{.5em} $^3$Toyota Technological Institute at Chicago \hspace{.5em} $^4$The Hebrew University of Jerusalem \\
{\tt\small \{wnhsu,talr,bshi,jdonley,adiyoss\}@meta.com}
}

\maketitle

%%%%%%%%% ABSTRACT
\begin{abstract}
Prior works on improving speech quality with visual input typically study each type of auditory distortion separately (e.g., separation, inpainting, video-to-speech) and present tailored algorithms. This paper proposes to unify these subjects and study Generalized Speech Enhancement, where the goal is not to reconstruct the exact reference clean signal, but to focus on improving certain aspects of speech. In particular, this paper concerns intelligibility, quality, and video synchronization. 
We cast the problem as audio-visual speech resynthesis, which is composed of two steps: pseudo audio-visual speech recognition (P-AVSR) and pseudo text-to-speech synthesis (P-TTS). P-AVSR and P-TTS are connected by discrete units derived from a self-supervised speech model. Moreover, we utilize self-supervised audio-visual speech model to initialize P-AVSR. The proposed model is coined ReVISE. 
ReVISE is the first high-quality model for in-the-wild video-to-speech synthesis and achieves superior performance on all LRS3 audio-visual enhancement tasks with a single model. 
To demonstrates its applicability in the real world, ReVISE is also evaluated on EasyCom, an audio-visual benchmark collected under challenging acoustic conditions with only 1.6 hours of training data. Similarly, ReVISE greatly suppresses noise and improves quality. Project page: \url{https://wnhsu.github.io/ReVISE}.
% Models and codes are available at \url{https://github.com/facebookresearch/av_hubert}.
\end{abstract}

%%%%%%%%% BODY TEXT
\section{Introduction}
\label{sec:intro}

Unlike anechoic studio recordings, speech in-the-wild is rarely clean: outdoor recordings are corrupted with all sorts of natural and non-natural sounds like wind and traffic noise~\cite{bello2019sonyc}. Speech recorded indoor often contains reverberation, mechanical noise, and overlapping speech from non-target speakers~\cite{watanabe2020chime}. On top of those, recording devices and network may also introduce other types of distortion, such as amplitude clipping, band-pass filtering, and package loss~\cite{adler2011audio}. Distortion makes it hard for both human and machines to comprehend speech~\cite{cooke2006glimpsing,lippmann1996speech}. Improving the quality and the intelligibility of corrupted speech is essential for assistive listening and robust speech processing. Generating clean speech signal based on its corrupted version is herein referred to as speech enhancement.

In speech enhancement, one line of research uses visual speech to provide auxiliary information~\cite{gabbay2017visual,gao2021visualvoice,ephrat2018looking,yang2022audio}, which is known as audio-visual speech enhancement. Audio-visual speech (e.g., talking-head videos) can be seen as a multimodal view of the speech. Since visual modality is immune to acoustic noise, combining both views enables more robust estimation of shared generating factors such as textual content.
Meanwhile, despite sharing the same goal of recovering corrupted speech, prior work often treats enhancement from each type of distortion as a separate problem: speech denoising and dereverberation addresses additive and convolutive non-speech noises~\cite{gabbay2017visual}, speech separation focuses on speech noises that exhibit similar characteristics to the target speech~\cite{gao2021visualvoice}, speech inpainting aims to recover dropped audio frames~\cite{morrone2021audio}, and video-to-speech synthesis is the extreme case of inpainting where all the frames are dropped~\cite{ephrat2017vid2speech, mira2022svts}. As a result, algorithms designed for one type of distortion may not be effective for another.

\begin{figure}[t]
    \centering
    \includegraphics[width=\linewidth]{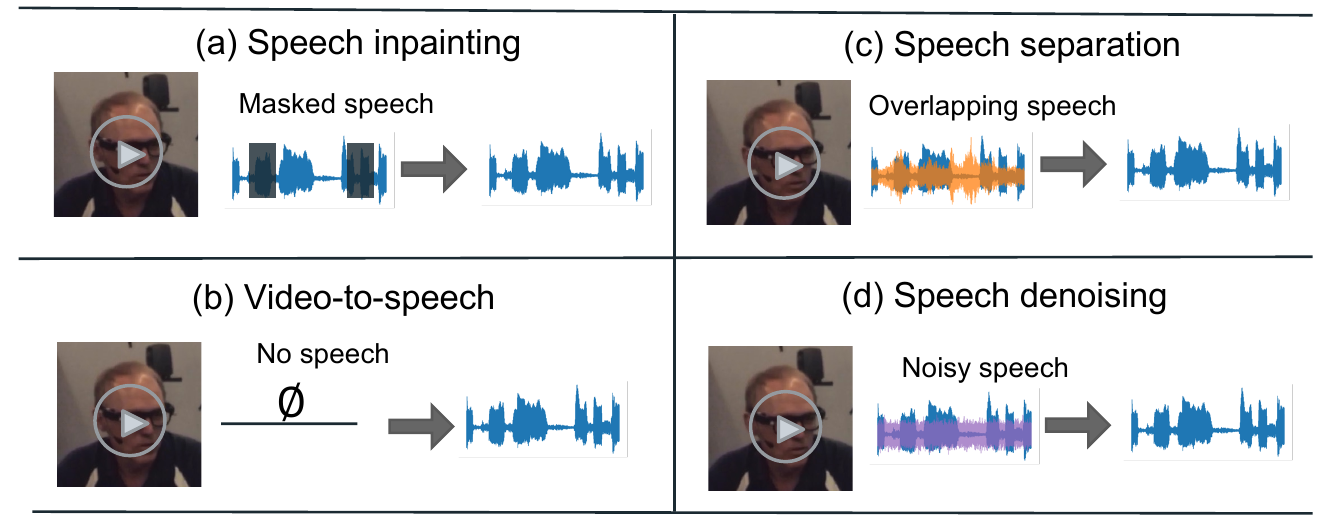}
    \caption{Illustration of AVSE with various distortion.}
    \label{fig:distortion}
\end{figure}

In this paper, we advocate a more holistic approach to audio-visual speech enhancement, where an algorithm should be evaluated on all types of distortion, and a single model should also be effective on all types of corrupted data, shifting from building distortion-specific models to a universal model. Following~\cite{serra2022universal}, we coin the concept \textbf{universal speech enhancement}.
In turn, we also argue that exact reconstruction of the reference clean speech is not an appropriate objective especially when the level of distortion is high, many different clean samples could have generated the same observed sample after distortion. 
To address the issue, we propose to relax the objective and solve the \textbf{generalized speech enhancement} (GSE) problem: instead of focusing on exact reconstruction and measuring metrics like signal-to-noise ratios (SNRs), the goal of GSE is to enhance a predefined set of attributes, such as content intelligibility, synchronization, and quality, which for example can be measured with word error rates (WERs), SyncNet scores~\cite{chung2016out}, and mean opinion scores (MOS). This leaves the models with the freedom to decide what to render for other aspects, such as voice and pitch contour, recovery of which are not essential for many applications.

This paper focuses on recovering intelligibility, synchronicity and quality. The task of improving those could be broken down into two steps: predicting the frame-level content and synthesizing high quality audio from it. 
Inspired by the resemblance to audio-visual speech recognition and speech synthesis, we propose ReVISE, short for \textbf{Re}synthesis with \textbf{V}isual \textbf{I}nput for \textbf{S}peech \textbf{E}nhancement. ReVISE is composed of a pseudo audio-visual speech recognition model (P-AVSR) and a pseudo text-to-speech synthesis model (P-TTS); instead of using text as the output/input of the two models, self-supervised speech units that encode speech content~\cite{hsu2021hubert,polyak2021speech} are adopted to bridge them, making the system free of text supervision. Furthermore, observing the gain on speech recognition brought by self-supervised learning, we also initialize the P-AVSR with a self-supervised audio-visual speech model, AV-HuBERT~\cite{shi2022learning}, which significantly improves the performance, especially on low-resource setups.

To demonstrate the universality and compare with the literature, we construct four types of corrupted speech using Lip-reading Sentences 3 (LRS3)~\cite{afouras2018lrs3} and AudioSet~\cite{gemmeke2017audio}, including audio-visual denoising, separation, inpainting, and video-to-speech. 
% The proposed system is evaluated on both distortion specific and agnostic setups. 
Results suggest that ReVISE is the first model capable of high-quality in-the-wild video-to-speech synthesis, while prior models fail to produce intelligible content~\cite{hassid2022more} or generate low-quality audio for in-the-wild videos~\cite{mira2022svts}. Compared to a strong masking-based method~\cite{gao2021visualvoice} on denoising and separation, ReVISE achieves comparable performance on mid-/high-SNR conditions (0-20dB), and are significantly stronger on lower SNR conditions, reducing WERs by up to 37.5\% absolute and improving MOS by up to 1.09. Finally, we also show that a single ReVISE model can tackle all four types of distortion with similar performance to distortion-specific models.
To further show the data efficiency and effectiveness of ReVISE on real data, we evaluate it on EasyCom~\cite{donley2021easycom}, an audio-visual speech dataset addressing the cocktail party problem which contains clean close-talking recordings and noisy distant recordings with background noise, loud interfering speech, and room reverberation. Results show that ReVISE still shines in this challenging setup, reducing WER by up to 32\% while other methods fail.

\section{Background}
\label{sec:related}
This section introduces the setup of audio-visual speech enhancement tasks and how prior studies approach these problems separately or jointly. We then contrast our approach with the literature and introduce self-supervised speech resynthesis --- the backbone of the proposed model.

\subsection{Audio-visual speech enhancement tasks}
In general, audio-visual speech enhancement is the task of improving the quality of corrupted speech $\tx_a$ given its corresponding talking head video $x_v$. During training, tuples of $(\tx_a, x_v, x_a)$ are provided where $x_a$ is the reference clean speech. The tasks are further divided depending on the type of distortion applied to $\tx_a$: speech denoising and dereverberation considers $\tx_a[t] = x_a[t] * h[t] + n[t]$ with impulse response $h$ and additive noise $n$ ($t$ indexes discrete time steps). Source separation considers $\tx_a[t] = x_a[t] + x'_a[t]$ where $x'_a$ is an interfering speech sampled from the same marginal distribution as $x_a$. For these two sub-tasks, masking-based methods (in magnitude spectrogram domain~\cite{narayanan2013ideal,erdogan2015phase}, complex spectrogram domain~\cite{ephrat2018looking, gao2021visualvoice, williamson2017time}, or on learned basis~\cite{luo2018tasnet, luo2019conv}) are widely adopted where the enhancement model $f$ predicts a mask $\hM = f(\tx_a, x_v)$. It is feasible because the target speech can be written as $X = M \cdot \tX$ where $X$ and $\tX$ are the clean and noisy speech in the target domain and $M$ is the ideal ratio mask. On the other hand, speech inpainting and video-to-speech synthesis considers $\tx_a[t] = m[t] \cdot x_a[t]$ where $m[t] \in \{0, 1\}$ for inpainting and $m[t] = 0$ for the latter. Masking-based methods are not feasible for these tasks. Hence, previous studies directly predict spectrogram~\cite{mira2022svts,morrone2021audio} or waveform~\cite{mira2022end,chazan2021single, nachmani2020voice} as $\hx_a = f(\tx_a, x_v)$ and optimize the model with regression and adversarial losses. Nevertheless, previous models have not been able to generate realistic samples with in-the-wild datasets like LRS3, because these models are deterministic while the underlying mapping is highly stochastic, especially when there is a large portion of frames dropped. 

To accommodate general distortions, universal enhancement models have to adopt generation-based method~\cite{serra2022universal,polyak2021high,pascual2019towards,su2021hifi,liu2021voicefixer}, otherwise they cannot handle distortions like package loss (i.e., inpainting).
For the same reason, our proposed model is also generation-based. However, it differs from previous works in two main aspects. First, prior universal enhancement models are audio-based, which do not leverage auxiliary input like visual speech. Hence, source separation and enhancement from silence (i.e., cross-modal generation) are not attainable and hence have not been included. Second, regardless of the training objective (regression~\cite{nair2021cascaded}, adversarial~\cite{pascual2019towards,su2021hifi,polyak2021high}, or score-matching~\cite{serra2022universal}), previous universal models predict reference clean speech directly. We consider a different paradigm based on self-supervised speech resynthesis~\cite{polyak2021speech}: our model predicts a self-supervised representation $z=k(x_a)$ of the reference clean speech. A separate model is used to convert $z$ back to audio, which is described in the next section.

\subsection{Self-supervised speech resynthesis}\label{sec:resyn}
Previous studies show that discretized SSL speech features from the Hidden Unit BERT (HuBERT) model~\cite{hsu2021hubert} encode mostly phonetic information and less about speaker and noise characteristics. HuBERT is pre-trained with a masked cluster prediction objective, where spans of input waveform are randomly masked and the model is asked to predict cluster assignment for those frames. The cluster assignment is iteratively refined: the initial one is obtained by clustering MFCC features with K-means, and subsequent ones are generated by clustering HuBERT features from the previous iteration. 
Unlike VQ-VAE~\cite{tjandra2020transformer}, HuBERT does not learn to encode all the factors; instead, it encodes information pertinent to inferring the unseen content, such as phonetic information. 
These properties make HuBERT units great candidates for replacing speech for text-free speech generative models, because they encode less nuisance variation and a model can predicts these units more easily compared to predicting speech directly~\cite{lakhotia2021generative, lee2021direct}. Such needs motivates the development of speech resynthesis models that convert units back to audio, because these generative models still have to produce speech as their final output.

\cite{polyak2021speech} proposed the first end-to-end model directly generating waveform from SSL units by adapting HiFi-GAN~\cite{kong2020hifi} which was originally designed as a vocoder converting spectrogram to waveform. The HiFi-GAN model is composed of a generator and a discriminator. The generator has a series of transposed convolution blocks with residual and dilated connections.
The discriminator comprises of multi-period and multi-scale sub-discriminators. 
The entire model is optimized with a combination of regression and adversarial losses.
The adpated HiFi-GAN takes SSL units, pitch, and speaker embedding as input, the latter two of which are added to capture variation of training data not encoded in the SSL units. Empirically, \cite{lee2021direct} demonstrates that speaker embedding and F0 can be removed if the model is trained on a high-quality single-speaker dataset such as LJSpeech~\cite{ljspeech17}.

\section{Method}
\label{sec:method}
In this section, we formulate the generalized speech enhancement problem formally, and introduce the proposed model in details.

\subsection{Problem Formulation}
\label{sec:gse}
Without loss of generality, we assume that original speech $x_a$ and its auxiliary view $x_v$ are generated with a bijective mapping $g_{m,Y}$ as $x_{m} = g_{m,Y}(\ys), m \in \{a, v\}$ given a set of factors $Y = \ys$ (e.g., pitch, speed, textual content) sampled from $p_Y$.
Corrupted speech $\tx_a$ is generated with a corruption function $g_d$ as $\tx_a = g_d(x_a, d)$ given $x_a$ and distortion parameter $d$ sampled from $p_d$. Audio-visual speech enhancement can be seen as estimating $p(Y \mid g_d(x_a, d),\ x_v)$. 
When the level of distortion is low, there are fewer $Y$ that can result in the same observed $\tx_a$ after corruption. Formulating it as a regression problem and measuring the performance of reconstructing \textit{all factors} with metrics like SNR is reasonable.
However, when the level of distortion is high, there exist multiple sets of $Y$ rendering the same noisy speech $\tx_a$. Take video-to-speech as an example: while the content and the timing of each word can be more accurately inferred, the exact pitch is hard to infer from the video.
In other words, $p(y_i | \tx_a,\ x_v)$ is more deterministic for some $y_i$ but not for the others. 

We argue that for these scenarios reconstructing the exact clean reference signal $x_a$ is an ill-posed problem. Instead, one should consider the generalized speech enhancement problem, where the goal is to generate an enhanced signal $\hx_a = f(\tx_a, x_v)$ that is on the manifold of clean speech $g_{a,Y}(\cdot)$ (high quality) and preserves the factors of interest (partial faithfulness). 
Let $g_{a,Y,i}^{-1}$ be the inverse mapping from speech to factor $y_i$, and $\ysp$ be the list of factors of interest (which should be a subset of factors that can be inferred from corrupted speech). Faithfulness is measured by the discrepancy between $g_{a,Y,i}^{-1}(\hx_a)$ and $g_{a,Y,i}^{-1}(x_a)$ for $i \in I^+$. 
In case of $y_i$ being the textual content, such discrepancy can be measured by the word error rate produced by an off-the-shelf speech recognizer for example.
Quality can be measured by metrics commonly used for text-to-speech synthesis such as mean opinion scores, where human raters rate how realistic and clean the audio sounds regardless of the textual content.
In this paper, we consider $\ysp$ to be textual content and timing (synchronization with the video) and do not address other factors such as voice, which is not of primary importance for communication.

\subsection{ReVISE}
\begin{figure}[t]
    \centering
    \includegraphics[width=\linewidth]{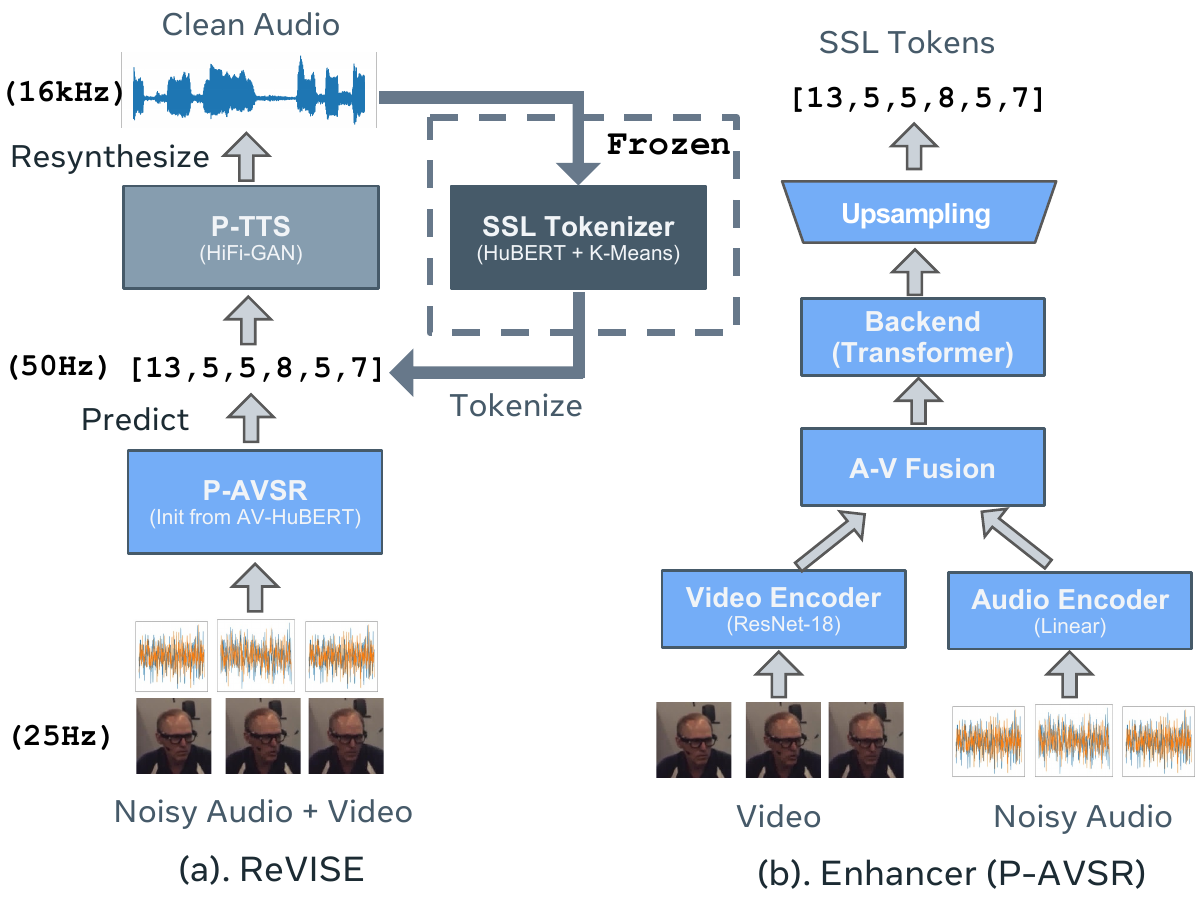}
    \caption{Model diagram.}
    \label{fig:revise}
\end{figure}

We propose to solve the problem in two stages: (i) predicting $\ysp$ of the clean reference $x_a$ given $(\tx_a, x_v)$, and (ii) resynthesizing speech conditioned on $\ysp$. The two steps are handled by a pseudo audio-visual speech recognition module (P-AVSR) and a pseudo text-to-speech synthesis module (P-TTS), respectively, as shown in \cref{fig:revise}. The discrete SSL units derived from HuBERT~\cite{hsu2021hubert} as described in \cref{sec:resyn} are used to represent $\ysp$. 
Using SSL units have two main benefits: not restricted by text availability and better encoding of non-verbal information.

The P-AVSR module performs a task very similar to speech recognition. Inspired by the recent success of self-supervised pre-training for audio~\cite{baevski2020wav2vec,hsu2021hubert} and audio-visual speech recognition~\cite{shi2022learning}, we initialize P-AVSR with AV-HuBERT~\cite{shi2022learning}, an SSL model achieving state-of-the-art performance on audio, visual, and audio-visual speech recognition. It is composed of a video encoder and an audio encoder as the modality-specific front-end, followed by a stack of transformer layers as the shared back-end taking concatenated audio and video features as input (\cref{fig:revise} right). The model encodes waveform into features at a frame rate of 25Hz, while the SSL unit from HuBERT are encoded at 50Hz. To match the rate, a lightweight transposed convolution layer followed by a softmax layer is added as the prediction head, which is randomly initialized.
The P-AVSR module is trained with a cross-entropy loss $L=\sum_t\sum_{j=1}^{C} z_t^j \log f_t^j(\tx_a, x_v)$, where $\mathbf{z}_t$ denotes the one-hot discrete unit label of the $t$-th frame, $f_t(\tx_a, x_v)$ is the predicted distribution over discrete units from the enhancer for the $t$-th frame and C unit vocabulary size.

We use the unit HiFi-GAN described in \cref{sec:resyn} for converting predicted $z$ into speech. Within ReVISE, the purpose of HiFi-GAN is to faithfully decode the units encoding target factors into speech. To prevent HiFi-GAN from introducing artifacts in the decoding process, we train the model on datasets with little nuisance variation such that $z$ to $x_a$ is more deterministic.

\section{Experimental Setup}
\label{sec:setup}

\subsection{Datasets} 
\label{sec:data}
We consider two datasets for our experiments. The first one is Lip Reading Sentences 3 (LRS3)~\cite{afouras2018lrs3}, which is a clean dataset based on TED talk videos. It contains 433 hours of audio-visual speech data and their corresponding text transcripts. Following prior work on speech enhancement, speech data are artificially corrupted to simulate enhancement tasks. Four tasks are created: (i) \textbf{Speech denoising:} in which we artificially mix a clean speech signal with a noise sample using randomly sampled Signal-to-Noise Ratio (SNR) in the range of [-20,20]. Noisy samples were selected from AudioSet~\cite{gemmeke2017audio}; (ii) \textbf{Speech source separation:} in which we follow a similar procedure to the speech denoising setup but instead of mixing the speech signal with a noisy sample, we mix it with another speech signal produced by a different speaker; (iii) \noindent \textbf{Speech inpaiting:} under this setup we randomly dropped (i.e., zero out) spans of $s\in\{20, 30, 40\}$ frames, with probability $p\in\{0.3, 0.4, 0.5\}$. Each frame corresponds to 20ms, hence in case $s=20$ we masked 400ms of audio. This process can be seen as a packet loss simulation; and (iv) \textbf{Video-to-speech synthesis:} in this setup we use the LRS3 dataset without any augmentations, where our goal is to synthesize the audio given a silent video only. 

The second dataset is the Easy Communications (EasyCom) dataset~\cite{donley2021easycom}, which is designed to study the cocktail party problem in the conversational augmented reality setup. In each session, one participant wears the glasses that record ego-centric video and 6-channel audio clips. Other participants wear a single-channel close-talking microphone. The dataset contains twelve 30 minute conversations, amounting to about six hours of raw recording. After keeping only speech segments and removing utterances without the target speaker being visible throughout the entire segment (using the official time-aligned transcripts and lip bounding boxes), there are 1.64/0.59/0.35 hours remained for the train/valid/test splits.\footnote{Session 4 and 12 are used for validation, and 10 and 11 for testing.}
The recordings from EasyCom and very different and much more challenging than those in LRS3. 
In EasyCom, there are motion blur in the video due to the head movement and barrel distortion from the camera. The distant microphone recordings often have interfering speech that is much louder than the target speech, coming from the subject who wears the glasses. The clips are recorded indoor with noise played by multiple speakers in the background. Hence, the microphone array on the glasses records substantial amount of noise and reverberation. We refer readers to the supplementary material to listen to the samples.

We use the close-talking microphone as the clean reference speech $x_a$ and distant one as the noisy speech $\tx_a$ following~\cite{donley2021easycom}. As a standard practice in multichannel speech processing, we consider beamformed audio that is derived from the four non-binaural channels as input~\cite{gannot2017consolidated,beiton2022audio,donley2021easycom}.
We use a maximally directive beamformer formulation that is optimized using a minimum-variance distortionless-response algorithm with a diffuse noise covariance and anechoic steering vector. The beamformer is steered towards the target's head location and the filter and sum is performed using weighted overlap add. We follow the exact formulation and implementation in~\cite{donley2021easycom}. We use a 64ms length FFT and filter length at 16kHz sampling rate, with an analysis and synthesis hanning window at 50\% overlap. The diffuse noise covariance and target steering vector are obtained using the set of array transfer functions (ATFs) provided in the dataset and the distortionless response reference is microphone number two.

For noisy speech $\tilde{x}_a$ used in training, we merge all six channels, beamformed audio as well as the audio from close-talking microphone to train ReVISE. Combining the multiple views of "same" speech can be regarded as one form of data augmentation, which has empirically improve performance a lot (see Appendix). For evaluation, we test our model on both channel two (i.e., single-channel) and beamformed audio (i.e., multi-channel).

\subsection{Model}
\paragraph{SSL speech tokenizer}
We use a \textsc{Base} HuBERT model which is composed of a convolutional encoder and 12 Transformer layers. Each layer has embedding dimension of 768, feed-forward layer dimension of 3072, and 12 self-attention heads. The model is pre-trained for three iterations on 32 GPUs with 400K updates per iteration, using clusters of MFCC/6-th layer feature from iteration 1/9-th layer feature from iteration 2 as the target, with a codebook size of 100/500/1000, respectively, following the recipe of \cite{lee2021direct}. A total of 221K hours of unlabeled speech data combining multilingual Librispeech~\cite{pratap2020mls}, Common Voice~\cite{ardila2019common}, and VoxPopuli~\cite{wang2021voxpopuli} in eight languages (En, Es, Fr, De, Nl, It, Pl, Pt) are used for pre-training. The final SSL units $z$ used in ReVISE are generated by clustering the third iteration feature at the last layer with a codebook size of 2000 with K-means.
% optional: add #gpus, batch size

\paragraph{P-TTS} We use the unit HiFi-GAN described in \cref{sec:resyn}. The model is trained for 400K updates on 8 GPUs on the LJSpeech~\cite{ljspeech17} dataset resampled to 16kHz to match the sample rate of other datasets. Similar to~\cite{lee2021direct}, F0 and speaker embedding are not used. 

\paragraph{P-AVSR}
We use the \textsc{Large} AV-HuBERT model by default from \cite{shi2022learning}. It takes video and/or audio as input, where video is head-crop image frames at 25fps, and audio is 23-dimensional Mel filterbank sequence (FBank) computed with a 10ms frame shift and stacked every four frames to match the video. The video encoder is a ResNet-18~\cite{ma2021end} model, while the audio-encoder is simply a linear projection. The features are concatenated frame-by-frame and passed to a stack of 24 Transformer layers, which has embedding and feed-forward layer dimension of 1024 and 4096 with 16 attention heads. For all but EasyCom experiments we use the publicly available checkpoints from \cite{shi2022learning} that were pre-trained on LRS3 and filtered VoxCeleb2 for initialization. For Easycom, a u-HuBERT model~\cite{hsu2022uhubert} pre-trained additionally with Librilight~\cite{kahn2020libri} is used for initialization, which improves upon using~\cite{shi2022learning}.
The up-sampling module is one transposed convolution layer with stride of 2, kernel size of 4, and 768 channels, followed by GeLU activation~\cite{hendrycks2016gaussian}. The P-AVSR models are fine-tuned on 8 GPUs for less than 45K updates. Detailed fine-tuning configurations for each task are presented in the appendix.

\subsection{Evaluation}
We evaluate the proposed model and baselines on the following axes: \textit{content}, \textit{synchronization}, \textit{quality}, and \textit{low-level detail} reconstruction. As discussed in \cref{sec:gse}, we focus on the first three axes, and the results for the last metric are included in the appendix for completeness.

For content, we use the WER computed with a speech recognition model to measure the intelligibility quantitatively similar to~\cite{mira2022svts, mira2022end}. The public model from \cite{xu2021self} is used, which reports a WER of 5.6\% on LRS3 test split and 35.7\% on the EasyCom close-talking validation set. For synchronization, following~\cite{hassid2022more} we use SyncNet~\cite{chung2016out} metrics, the predicted temporal distance between audio and video (LSE-D) and the prediction's confidence (LSE-C) are averaged over the entire test set. For quality, we follow the tradition of text-to-speech synthesis evaluation and conduct subjective mean opinion score (MOS) studies with a scale from 1 to 5 and a 0.5 increment. We evaluate 50 randomly sampled files from the test set, where each sample was evaluated by at least 15 raters using the CrowdMOS package~\cite{ribeiro2011crowdmos}. Finally, to evaluate reconstruction of low-level details as typically done in speech denoising or source separation studies, we include ESTOI~\cite{taal2011algorithm} and Mel cepstral distortion (MCD)~\cite{kubichek1993mel}. While MCD provides an estimate of the similarity of two signals by measuring the mel cepstra difference, ESTOI put more emphasis on speech properties. 

\section{Results}
\label{sec:result}

Quantitative results are presented in the section. We strongly recommend readers to watch the samples provided in the supplementary material for better understanding of the sample quality compared to the baselines.

\subsection{Ground truth and resynthesis performance}
For reference, we first evaluate reference clean speech (Tgt. audio) and resynthesized clean speech (Tgt. audio resynthesis) with the proposed metrics and report the results in \cref{tab:lrs3_v2a}. The gap between the two is an approximation of how much information is lost when tokenizing speech into SSL units. It shows that intelligibility, synchronization, and quality are all slightly degraded.
As ReVISE is trained to predict the tokens inferred from clean speech, if the ReVISE model have 100\% prediction accuracy then the performance will be identical to the ``Tgt. audio resynthesis'' row here. Hence, the performance on the resynthesized speech roughly upper-bounds that of the ReVISE model.

\begin{table}[t]
  \centering
  \resizebox{\linewidth}{!}{
  \begin{tabular}{@{}l|c|cc|c@{}}
    \toprule
    & Cont & \multicolumn{2}{c|}{Sync} & Qual \\
    Method & WER$\downarrow$ & LSE-C$\uparrow$ & LSE-D$\downarrow$ & MOS$\uparrow$ \\
    \midrule\midrule
    Tgt. audio & 5.6\% & 5.20 & 6.33 & 4.38\pmr{0.02} \\
    Tgt. audio resynthesis & 10.6\%	& 5.02 & 6.40 & 4.16\pmr{0.02} \\
    \midrule
    \multicolumn{5}{@{}l}{\textit{Silent video to speech}} \\
    \rowcolor{gray!10} SVTS-L, LRS3~\cite{mira2022svts} & 81.5\% & 5.07 & 6.46 & 2.12\pmr{0.04} \\
    \rowcolor{gray!10} SVTS-L, LRS3+VC2~\cite{mira2022svts} & 67.4\% & 5.53 & 6.00 & 2.23\pmr{0.03} \\
    ReVISE (Ours) & 33.9\% & 5.03 & 6.44 & 4.13\pmr{0.02} \\
    \bottomrule    
  \end{tabular}
  }
  \caption{Results for video-to-speech synthesis task. We report WER, LSE, and MOS scores with confidence interval at 95\%. Results are reported on LRS3 test set.}
  \label{tab:lrs3_v2a}
\end{table}

\subsection{Video-to-speech synthesis}

In \cref{tab:lrs3_v2a}, we compare ReVISE with SVTS~\cite{mira2022svts},\footnote{We obtained test set samples of SVTS from the authors for evaluation.} which is the state-of-the-art model on the video-to-speech synthesis task. SVTS is composed of a video-to-spectrogram predictor and a separately trained neural vocoder. Its video-to-spectrogram predictor is trained with a regression loss, which takes as input a lip video and a speaker embedding extracted from a pre-trained fixed speaker encoder. Speaker embedding is needed because voice cannot be accurately inferred from the video. The model demonstrate strong results on two constrained datasets (LRW~\cite{chung2016lip} and GRID~\cite{cooke2006audio}) which contain isolated words and fixed-length sentences with limited word choices, respectively. 

As shown in \cref{tab:lrs3_v2a}, because the audio quality generated by SVTS is mediocre (2.12 and 2.23 MOS),\footnote{\url{https://sites.google.com/view/scalable-vts}} these samples are not very intelligible by machines, reflected by the high WER (81.5\% and 67.4\% when trained on LRS3 and LRS3+VoxCeleb2).
In contrast, our proposed model generates much higher quality audio (4.13 MOS) and yields significantly lower WER (33.9\%). 
Through manual inspection, we found a confidence score above 5 and distance below 6 implies good synchronization on LRS3. All methods generate audio fairly synchronized to the video.

\begin{table}
  \centering
  \resizebox{.85\linewidth}{!}{
  \begin{tabular}{@{}lcc|c|cc@{}}
    \toprule
    & & & Cont & \multicolumn{2}{c}{Sync} \\
    Method & Mod & Test & WER$\downarrow$ & LSE-C$\uparrow$ & LSE-D$\downarrow$ \\
    \midrule\midrule

    \multicolumn{5}{@{}l}{\textit{Audio-visual speech inpainting}} \\
    
    \rowcolor{gray!10} && 30\% & 44.1\% & 4.37 & 7.05 \\
    \rowcolor{gray!10} && 40\% & 53.5\% & 4.01 & 7.36 \\
    \rowcolor{gray!10} \multirow{-3}{*}{Inp. audio} & \multirow{-3}{*}{A} 
                       & 50\% & 60.3\% & 3.78 & 7.60 \\
    % \midrule
                       
    \rowcolor{gray!0} && 30\% & 51.2\% & 4.20 & 7.14 \\
    \rowcolor{gray!0} && 40\% & 60.1\% & 3.94 & 7.38 \\
    \rowcolor{gray!0} \multirow{-3}{*}{Resynthesis} & \multirow{-3}{*}{A} 
                       & 50\% & 66.3\% & 3.68 & 7.63 \\

    \rowcolor{gray!10} && 30\% & 43.9\% & 4.37 & 7.04 \\
    \rowcolor{gray!10} && 40\% & 53.5\% & 4.04 & 7.34 \\
    \rowcolor{gray!10} \multirow{-3}{*}{Demucs~\cite{defossez2020real}} & \multirow{-3}{*}{A} 
                        & 50\% & 60.0\% & 3.82 & 7.57 \\

    \rowcolor{gray!0} && 30\% & 14.9\% & 5.01 & 6.42 \\
    \rowcolor{gray!0} && 40\% & 18.1\% & 4.99 & 6.44 \\
    \rowcolor{gray!0} \multirow{-3}{*}{ReVISE (Ours)} & \multirow{-3}{*}{AV} 
                        & 50\% & 19.2\% & 5.04 & 6.41 \\
      
    \bottomrule
  \end{tabular}
  }
  \caption{Results of speech inpainting on LRS3 test set with 30\%/40\%/50\% of frames dropped.}
  \label{tab:inpaint}
\end{table}

\subsection{Audio-visual speech inpainting}
We next evaluate ReVISE on the audio-visual speech inpainting task with three test splits, where 30\%, 40\%, and 50\% of the frames are dropped with dropped spans longer than 400ms. While there have been several works studying audio-based speech inpainting, \cite{morrone2021audio} is the only work we are aware of studying the audio-visual setup. However, their model was only evaluated on the very constrained GRID dataset, showing a phone error rate of 13.7\%. SVTS achieves a word error rate of 17.9\%, hinting that \cite{morrone2021audio} is likely worse than SVTS even when having partial audio as input. Accordingly, we decide not to compare with \cite{morrone2021audio} but compare with a publicly available strong generation-based audio enhancement model, Demucs~\cite{defossez2020real}, which is trained on $\sim$650 hours of high quality clean speech mixed with noise from AudioSet~\cite{gemmeke2017audio} and FreeSound\footnote{\url{https://freesound.org}}. In addition, we consider resynthesis as another baseline, which encodes and decodes the corrupted speech with the same SSL tokenizer and P-TTS model.

``Inp. audio'' row presents the results evaluated on the corrupted audio in \cref{tab:inpaint}. We see that intelligibility degrades (WER increases from 5.6\% to 60.3\%) as the percentage of dropped frames increases, and so does synchronization (5.20 to 3.78 for confidence and 6.33 to 7.60 for distance). The two audio-based baselines are not effective, showing similar or even worse results compared to the corrupted samples before enhancement. 
The little gain from Demucs over the inputting-audio baseline shows that existing enhancement approaches fails to generalize to inpainting.
In contrast, our proposed model is effective on improving both intelligibility and synchronization. The WER drops from 60.3\% to 19.2\% after enhancement when half of the audio is missing in the corrupted audio. Comparing enhancing from no audio at all (video-to-speech synthesis) with enhancing from observing half of the audio (inpainting, 50\%), we observe a 14.7\% WER reduction (33.9\% $\rightarrow$ 19.2\%), showing that the model use the additional audio information effectively to improve content reconstruction. As for synchronization, ReVISE restore a similar level regardless of the percentage dropped.

\subsection{Audio-visual speech denoising}
\label{sec:denoise}

\begin{table}
  \centering
  \resizebox{\linewidth}{!}{
  \begin{tabular}{@{}lcc|c|cc|c@{}}
    \toprule
    & & & Cont & \multicolumn{2}{c|}{Sync} & Qual \\
    Method & Mod & Test & WER$\downarrow$ & LSE-C$\uparrow$ & LSE-D$\downarrow$ & MOS$\uparrow$ \\
    \midrule
    \rowcolor{gray!0}Tgt. audio & - & - & 5.6\% & 5.20 & 6.33 & 4.36\pmr{0.018}\\
    \midrule\midrule
    \multicolumn{5}{@{}l}{\textit{Audio-visual speech denoising}} \\
    \rowcolor{gray!10} && lvl 1 &  7.8\% & 4.66 & 6.77 & 2.95\pmr{0.03}\\
    \rowcolor{gray!10} && lvl 2 & 20.0\% & 3.80 & 7.52 & 2.82\pmr{0.03} \\
    \rowcolor{gray!10} && lvl 3 & 63.9\% & 2.41 & 8.68 & 2.79\pmr{0.02}\\
    \rowcolor{gray!10} \multirow{-4}{*}{Inp. audio} & \multirow{-4}{*}{A}
                       & lvl 4 & 87.7\% & 1.54 & 9.39 & 2.71\pmr{0.04} \\
                       
    \rowcolor{gray!0} && lvl 1 & 13.3\% & 4.98 & 6.46 & 4.33\pmr{0.02}\\
    \rowcolor{gray!0} && lvl 2 & 25.9\% & 4.78 & 6.64 & 4.37\pmr{0.02} \\
    \rowcolor{gray!0} && lvl 3 & 69.4\% & 3.64 & 7.61 & 3.63\pmr{0.02}\\
    \rowcolor{gray!0} \multirow{-4}{*}{Resynthesis} & \multirow{-4}{*}{A}
                       & lvl 4 & 90.8\% & 2.11 & 8.85 & 3.34\pmr{0.03} \\
    
    \rowcolor{gray!10} && lvl 1 & 6.9\%  & 5.15 & 6.38 &3.58\pmr{0.04}\\
    \rowcolor{gray!10} && lvl 2 & 15.1\% & 4.95 & 6.57 &  3.63\pmr{0.03}  \\
    \rowcolor{gray!10} && lvl 3 & 48.0\% & 4.14 & 7.28 & 3.36\pmr{0.03}\\
    \rowcolor{gray!10} \multirow{-4}{*}{Demucs~\cite{defossez2020real}} & \multirow{-4}{*}{A} & lvl 4 & 81.3\% & 2.43 & 8.75 & 2.89\pmr{0.04}  \\
    
    \rowcolor{gray!0} && lvl 1  & 6.6\% & 5.16 & 6.35 & 4.11\pmr{0.03}\\
    \rowcolor{gray!0} && lvl 2 & 8.8\% & 5.19 & 6.35  & 3.94\pmr{0.02}  \\
    \rowcolor{gray!0} && lvl 3 & 23.4\% & 5.07 & 6.45 & 3.80\pmr{0.02}\\
    \rowcolor{gray!0} \multirow{-4}{*}{VisualVoice~\cite{gao2021visualvoice}} & \multirow{-4}{*}{AV}
                        & lvl 4 & 58.0\% & 4.53 & 6.86 & 3.25\pmr{0.03} \\
    
    \rowcolor{gray!10} && lvl 1 & 9.4\%  & 5.03 & 6.42 & 4.31\pmr{0.02}\\
    \rowcolor{gray!10} && lvl 2 & 9.7\%  & 5.04 & 6.41 & 4.36\pmr{0.02} \\
    \rowcolor{gray!10} && lvl 3 & 11.7\% & 5.04 & 6.41 & 4.41\pmr{0.02}\\
    \rowcolor{gray!10} \multirow{-4}{*}{ReVISE (Ours)} & \multirow{-4}{*}{AV} 
                       & lvl 4 & 20.5\% & 5.07 & 6.40 & 4.34\pmr{0.02} \\
                       
    \bottomrule
  \end{tabular}
  }
  \caption{Results of speech deonoising on LRS3 test set. SNR range for lvl 1/2/3/4 are [10,20]/[0,10]/[-10,0]/[-20,-10] dB.}
  \label{tab:denoise}
\end{table}

\cref{tab:denoise} presents results of audio-visual speech denoising on four test splits: lvl 1-4, which contain noisy samples of SNR in [10, 20], [0, 10], [-10, 0], and [-20, -10]. 
In addition to the two audio-based baselines from the previous section, the proposed model is also compared with VisualVoice~\cite{gao2021visualvoice} in this and the next section, which is a state-of-the-art model for audio-visual source separation and denoising. VisualVoice predicts a complex IRM given noisy audio, lip video, and speaker face image as input and the model is optimized with a combination of mask prediction loss, cross-modal matching loss (image and enhanced audio embeddings from the same speaker should be close) and consistency loss (embeddings of enhanced audio for the same speaker should be close). To have a fair comparison, we re-train VisualVoice models with the same dataset as ReVISE using a similar setup provided by the authors.

In terms of intelligibility, we observe that the performance of resynthesizing corrupted audio is still bad. Demucs performs well for high SNR setup (lvl 1) but lags behind audio-visual models for the rest setups. VisualVoice achieves the best results on the [0, 20] dB SNR range, but ReVISE is much more robust to even a higher level of noise, reporting a WER of 20.5\% at lvl 4. Synchronization scores follow a similar trend to WERs in terms of ranking across models and test splits. Quality of the proposed model is significantly higher (4.31-4.41 MOS) than all baselines, and is on par with the reference clean speech (4.36 MOS).

\subsection{Audio-visual source separation}

\begin{table}
  \centering
  \resizebox{0.9\linewidth}{!}{
  \begin{tabular}{@{}lrr|r|rr@{}}
    \toprule
    & & & Cont & \multicolumn{2}{c}{Sync} \\
    Method & Mod & Test & WER$\downarrow$ & LSE-C$\uparrow$ & LSE-D$\downarrow$ \\
    \midrule\midrule
    \multicolumn{5}{@{}l}{\textit{Audio-visual speech separation}} \\
    
    \rowcolor{gray!10} && lvl 1 & 15.7\% & 4.73 & 6.74 \\
    \rowcolor{gray!10} && lvl 2 & 65.7\% & 3.85 & 7.48\\
    \rowcolor{gray!10} && lvl 3 & 102.3\% & 2.93 & 8.27 \\
    \rowcolor{gray!10} \multirow{-4}{*}{Inp. audio} & \multirow{-4}{*}{A}
                        & lvl 4 & 105.5\% & 2.46 & 8.66 \\
    
    \rowcolor{gray!0} && lvl 1 & 11.7  & 5.06 & 6.46 \\
    \rowcolor{gray!0} && lvl 2 & 55.7  & 4.10 & 7.25 \\
    \rowcolor{gray!0} && lvl 3 & 101.2 & 2.77 & 8.42 \\
    \rowcolor{gray!0} \multirow{-4}{*}{Demucs~\cite{defossez2020real}} & \multirow{-4}{*}{A}
                        & lvl 4 & 106.2 & 2.15 & 8.95 \\
    
    \rowcolor{gray!10} && lvl 1 &  6.2\% & 5.13 & 6.37 \\
    \rowcolor{gray!10} && lvl 2 &  7.8\% & 5.12 & 6.39 \\
    \rowcolor{gray!10} && lvl 3 & 15.3\% & 4.91 & 6.57 \\
    \rowcolor{gray!10} \multirow{-4}{*}{VisualVoice~\cite{gao2021visualvoice}} & \multirow{-4}{*}{AV}
                       & lvl 4 & 51.8\% & 4.20 & 7.17 \\
    
    \rowcolor{gray!0} && lvl 1 &  9.9\% & 5.05 & 6.41 \\
    \rowcolor{gray!0} && lvl 2 & 10.2\% & 5.05 & 6.41 \\
    \rowcolor{gray!0} && lvl 3 & 11.4\% & 5.05 & 6.41 \\
    \rowcolor{gray!0} \multirow{-4}{*}{ReVISE (Ours)} & \multirow{-4}{*}{AV} 
                        & lvl 4 & 15.7\% & 5.04 & 6.41 \\
    \bottomrule
    
  \end{tabular}
  }
  \caption{Results of source separation on LRS3 test set. SNR range for lvl 1/2/3/4 are [10,20]/[0,10]/[-10,0]/[-20,-10] dB.}
  \label{tab:sep}
\end{table}

Following \cref{sec:denoise}, four noisy test sets are created with different levels of SNR. Comparing the speech source separation results in \cref{tab:sep} and denoising results in \cref{tab:denoise}, we observe that audio model (Demucs) performs worse with speech noise than with non-speech noise at the same SNR level. In contrast, audio-visual models report similar performance on the two tasks at high SNR, and better performance on separation at low SNR. These results suggest that it is easier to remove speech noise if the model have auxiliary information that it can use to identify the target speech. Comparing VisualVoice and ReVISE, it reveals a similar trend where ReVISE is still substantially better at low SNR.

\subsection{Universal audio-visual speech enhancement}
\begin{table*}[t]
  \centering
  \resizebox{\linewidth}{!}{
  \begin{tabular}{@{}l|c|ccc|cccc|cccc|c@{}}
    \toprule
    & video-to-speech & \multicolumn{3}{c|}{speech inpainting WER} & \multicolumn{4}{c|}{speech denoising WER } & \multicolumn{4}{c|}{source separation WER} & Avg. \\
    & WER & 20\% & 30\% & 40\% & lvl 1 & lvl 2 & lvl 3 & lvl 4 & lvl 1 & lvl 2 & lvl 3 & lvl 4 & \\
    \midrule
    Separate ReVISE & 33.9\% & 14.9\% & 18.1\% & 19.2\% & 9.4\% & 9.7\% & 11.7\% & 20.5\% & 9.9\% & 10.2\% & 11.4\% & 15.7\% & 19.0\% \\
    Universal ReVISE & 34.5\% & 14.9\% & 17.9\% & 19.2\% & 8.7\% & 9.4\% & 11.5\% & 20.3\% & 9.1\% & 9.7\% & 10.9\% & 15.5\% & 18.9\% \\
    \bottomrule
  \end{tabular}
  }
  \caption{Comparing corruption-specific ReVISE with universal ReVISE on intelligibility measured by WER (\%).}
  \label{tab:universal}
\end{table*}
In this section, we build a single ReVISE model trained on all four types of distortion and compare it with corruption-specific models presented in earlier sections (\cref{tab:universal}) in terms of WER. Surprisingly, the universal model beats or matches the distortion-specific model on almost all tasks. The only exception is video-to-speech synthesis, on which the universal model is 0.5\% WER worse.

\subsection{AV enhancement on real data --- EasyCom}

\begin{table}
  \centering
  \resizebox{\linewidth}{!}{
  \begin{tabular}{@{}lr|cc@{}}
    \toprule
    Method & Mod & \shortstack[c]{WER$\downarrow$ (ch2) \\ (dev/test)} & \shortstack[c]{WER$\downarrow$ (bf)\\ (dev/test)} \\
    \midrule\midrule
    Tgt. audio & A & \multicolumn{2}{c}{35.7\% / 37.6\%} \\
    Tgt. audio resynthesis & A & \multicolumn{2}{c}{43.2\% / 47.7\%} \\
    \midrule
    \multicolumn{4}{@{}l}{\textit{Audio-visual speech enhancement}} \\
    \rowcolor{gray!10} \multirow{-1}{*}{Inp. audio} & A & 
        74.2\% / 87.5\% & 60.7\% / 71.5\% \\
    \rowcolor{gray!0} \multirow{-1}{*}{Demucs~\cite{defossez2020real}} & A & 
        76.9\% / 86.8\% & 63.4\% / 69.8\% \\
    \rowcolor{gray!10} \multirow{-1}{*}{Resynthesis} & A & 
        79.6\% / 91.0\% & 68.7\% / 77.6\% \\
    \rowcolor{gray!0} \multirow{-1}{*}{ReVISE (Ours)} & AV & 50.3\% / 55.0\% & 47.6\% / 52.1\% \\
    \bottomrule
    
  \end{tabular}
  }
  \caption{Audio-visual speech enhancement results on EasyCom (ch2: channel 2, bf: beamforming). Synchronization results are not reported because SyncNet fails even on clean reference $x_a$.}
  \label{tab:easycom-result}
\end{table}

We next study how effective ReVISE on EasyCom, a real noisy dataset that is much more challenging than LRS3 and has much fewer hours of training data. \cref{tab:easycom-result} compares ReVISE with several baseline methods enhancing single channel speech (ch2) or beamformed speech (bf) from multi-channel sources. Note that we attempted to re-train VisualVoice on EasyCom but were not successful on getting any meaningful results, likely due to the lack of data as well as the difficulty. In addition, we use a P-AVSR model that takes lip crops instead of head crops as input for this dataset, which provides slightly better performance. Ablation studies are included in the Appendix.

ReVISE substantially enhances the intelligibility of noisy speech in both setups (WER: 87.5\% $\rightarrow$ 55.0\% on ch2 and 71.5\% $\rightarrow$ 52.1\% on bf), while Demucs only managed to reduce the WER by 0.7\% and 1.7\%, respectively. Moreover, the WER of the enhanced beamformed speech from ReVISE is only 4.4\% away from resynthesized clean speech: the target ReVISE is trained to predict. One can expect the gap to clean speech (14.5\% WER) will be reduced if the quality of the SSL units improves.

\cref{tab:easycom-mos} presents the subjective quality evaluation results for EasyCom. We first note that the quality of resynthesized target audio (4.13 MOS) is in fact higher than the target audio itself (3.76 MOS). This is because the P-TTS module is trained on the high-quality LJSpeech, which has better quality than the EasyCom audio recorded by the close-talking microphones, as the latter still contain mild background noise and overlapping speech. 
Similarly, the predicted audio from the proposed ReVISE model also delivers similar levels of audio quality (4.19 and 4.11 with ch2 and beamformed input, respectively), which are significantly better than the original distant recordings (2.95 and 2.67) and the baseline methods (2.95 and 2.39 from Demucs). 

The results suggests a crucial advantage of the proposed ReVISE model compared to previous studies: for ReVISE, the audio quality is not bounded by the enhancement training data (LRS3 and EasyCom), but determined by the quality of P-TTS training data. In contrast, prior works such as SVTS or VisualVoice could only produce audio that is as good as the enhancement target.

\begin{table}[t]
  \centering
  \resizebox{\linewidth}{!}{
  \begin{tabular}{@{}lr|cc@{}}
    \toprule
    Method & Mod & \shortstack[c]{MOS$\uparrow$ (ch2)} & \shortstack[c]{MOS$\uparrow$ (bf)} \\
    \midrule\midrule
    Tgt. audio & A & \multicolumn{2}{c}{3.76\pmr{0.04}} \\
    Tgt. audio resynthesis & A & \multicolumn{2}{c}{4.13\pmr{0.03}} \\
    \midrule
    \multicolumn{4}{@{}l}{\textit{Audio-visual speech enhancement}} \\
    \rowcolor{gray!10} \multirow{-1}{*}{Inp. audio} & A & 
        2.95\pmr{0.04} & 2.67\pmr{0.03} \\
    \rowcolor{gray!0} \multirow{-1}{*}{Demucs~\cite{defossez2020real}} & A & 
        2.95\pmr{0.04} & 2.39\pmr{0.03} \\
    \rowcolor{gray!10} \multirow{-1}{*}{Resynthesis} & A & 
        3.97\pmr{0.04} & 3.98\pmr{0.03} \\
    \rowcolor{gray!0} \multirow{-1}{*}{ReVISE (Ours)} & AV & 
        4.19\pmr{0.02} & 4.11\pmr{0.04} \\
    \bottomrule
    
  \end{tabular}
  }
  \caption{Subjective quality evaluation of audio-visual speech enhancement results on EasyCom test set (ch2: channel 2, bf: beamforming).
  }
  \label{tab:easycom-mos}
\end{table}

\subsection{Ablation studies}

\begin{table}
  \centering
  \resizebox{\linewidth}{!}{
  \begin{tabular}{@{}l|cc@{}}
    \toprule
    % & \multicolumn{2}{c}{WER} \\
    Unit & Resyn WER$\downarrow$ & V2S WER$\downarrow$ \\
    \midrule\midrule
    AV-HuBERT (LRS3+VC2)~\cite{shi2022learning} & 20.7 & 41.2 \\
    HuBERT (LS960)~\cite{hsu2021hubert}     & 17.4 & 35.2 \\
    HuBERT (VP+MLS+CV)            & 10.6 & 33.9 \\
    \bottomrule    
  \end{tabular}
  }
  \caption{Ablation studies on SSL unit choices. Clean audio resynthesis (Resyn) and video-to-speech (V2S) 
  WERs are reported.}
  \label{tab:abl-unit}
\end{table}

\begin{table}
    \begin{minipage}[t]{0.38\linewidth}
    \centering
    \resizebox{\linewidth}{!}{
    \begin{tabular}{@{}l|cc@{}}
        \toprule
        P-AVSR & \multicolumn{2}{c}{V2S WER$\downarrow$}  \\
        Size & scratch & PT \\
        \midrule\midrule
        \textsc{Base}  & 81.0 & 42.9 \\
        \textsc{Large} & 78.6 & 35.5 \\
        \bottomrule    
    \end{tabular}
    }
    \end{minipage}%
    \hfill
    \begin{minipage}[t]{0.58\linewidth}
    \centering
    \resizebox{.95\linewidth}{!}{
    \begin{tabular}{@{}l|cccc@{}}
        \toprule
        & \multicolumn{4}{c}{WER$\downarrow$} \\
        Mod & Enh & Sep & Inpaint & V2S \\
        \midrule\midrule
        AV & 12.6 & 11.4 & 17.4 & 34.8 \\
        A  & 29.2 & 39.8 & 44.6 & n/a  \\
        \bottomrule    
    \end{tabular}
    }
    \end{minipage}%
    \caption{(Left) Ablation studies on model sizes and pre-training. Performance on video-to-speech is reported. (Right) Comparison of ReVISE and its audio-only counterpart.}
    \label{tab:abl-pt-size-mod}    
\end{table}

We conduct ablation studies to evaluate the importance of pre-training P-AVSR, P-AVSR model size, selection of audio SSL units, and visual input. WER are reported here because intelligibility exhibits the highest variation across models reported earlier.
\cref{tab:abl-unit} compares different SSL speech units on two tasks: clean speech resynthesis and video-to-speech generation. The two alternative types of units are derived from the \textsc{Base} AV-HuBERT model trained on LRS3+VoxCeleb2~\cite{shi2022learning} and the \textsc{Base} HuBERT model trained on Librispeech~\cite{hsu2021hubert}. The corresponding P-TTS model are also trained on the same LJSpeech dataset. We see that performance on the two tasks are correlated, where units that loss less content information (lower Resyn WER) are better prediction targets for the P-AVSR model (lower V2S WER). \cref{tab:abl-pt-size-mod} (Left) shows pre-training brings significant gains (38.9\% for \textsc{Base} and 43.1\% for \textsc{Large}) and \textsc{Large} model performs better with a 7.4\% WER reduction.

Finally, we study the importance of visual input by comparing an audio-visual ReVISE with an audio-only version of it. We consider the universal speech enhancement model setup but remove video-to-speech data as it is incompatible to the audio-only model.
\cref{tab:abl-pt-size-mod} (Right) presents aggregated results averaged over testing splits for each task. We observe that the model using both audio and video as input outperform the counterpart consistently with a large margin. Specifically, the level of degradation varies substantially across task: enhancement is least effected (16.6\%), followed by inpainting (27.2\%) and separation (28.4\%) --- this shows that models rely heavily on visual information to determine the target speaker as well as infill the missing information from dropped frames. 
Last but not least, we note that the audio-visual ReVISE perform decently on video-to-speech synthesis despite not being trained on the task at all, suggesting zero-shot generalization to unseen types of distortion.

\section{Conclusion}
\label{sec:conclusion}
This paper presents ReVISE, a novel paradigm for audio-visual speech enhancement that is universal and benefits greatly from the recent advances in self-supervised speech pre-training. Empirical studies demonstrate effectiveness of ReVISE on a popular benchmark with synthetic noise (LRS3) as well as on a conversational benchmark with real noisy data collected from extremely challenging acoustic conditions (EasyCom). In particular, audio enhanced by ReVISE is significantly more intelligible and of higher quality compared to those from the previous work. 

While this paper studies universal enhancement, it only concerns audio but not video distortion. For future work, we hope to study more general multimodal audio-visual speech enhancement where an enhancement model can recover distortion in both modalities.

\section*{Acknowledgement}
The authors thank Rodrigo Schonburg for sharing SVTS samples for evaluations, thank Ruohan Gao for the assistance on VisualVoice code, and thank Stavros Petridis and Anurag Kumar for helpful discussions.

% \clearpage
%%%%%%%%% REFERENCES
{\small
\bibliographystyle{ieee_fullname}
\bibliography{egbib}
}

\clearpage
\appendix
\section{Video samples}
A webpage showing the video samples from EasyCom and from the four tasks on LRS3 (video-to-speech, audio-visual speech inpainting, audio-visual speech denoising, and audio-visuak source separation) are included in the supplementary material. Readers are highly recommended to watch the samples to better understand how the proposed model perform compared to the baselines.

\section{Additional Results}

\subsection{EasyCom visual features and training data}
Table~\ref{tab:easycom-analysis} shows the impact of fine-tuning data and visual inputs on the model performance in EasyCom.

\textbf{Data} For our main results, the pre-trained ReVISE model is fine-tuned on EasyCom only (1.6 hours). Though merging other audio-visual datasets and EasyCom can increase the size of training data by orders of magnitude, we do not observe gain brought by such practice (row (a) vs. row (b) in Table~\ref{tab:easycom-analysis}). This is potentially due to the severe domain mismatch in two datasets on multiple aspects such as types of audio (rehearsed speech vs. conversation) and noise (simulated vs. natural). 

\textbf{Input} We also notice that using mouth regions as input is more effective in enhancing speech compared to directly feeding talking head into the model (row (a) vs. row (c) in Table~\ref{tab:easycom-analysis}). Mouth cropping helps remove unrelated visual background, thus bridging gap in visual domain between pretraining and fine-tuning. 

\begin{table}[h]
  \centering
  \begin{tabular}{@{}ll|cc@{}}
    \toprule
    Input & Data & WER (ch2) & WER (bf) \\
    \midrule\midrule
    (a). Mouth & EasyCom  & \textbf{50.3}\% & \textbf{47.6}\% \\
    (b). Mouth & EasyCom+LRS3 & 54.1\% & 49.3\% \\
    (c). Head & EasyCom & 56.2\% & 51.4\% \\
    \bottomrule    
  \end{tabular}
  \caption{Impact of visual input and fine-tuning data on speech enhancement in EasyCom. Numbers are on development set.}
  \label{tab:easycom-analysis}
\end{table}

\subsection{Predicting units versus spectrogram}

To perform video-to-speech synthesis, ReVISE differs from SVTS~\cite{mira2022svts} in two main aspects. First, we predict SSL units while SVTS predicts Mel spectrogram. Second, we initialize the video-to-unit prediction module (P-AVSR) with AV-HuBERT, a self-supervised audio-visual speech model while SVTS trains the video-to-spectrogram model from scratch. 

\begin{figure}[htp]
    \centering
    \includegraphics[width=\linewidth]{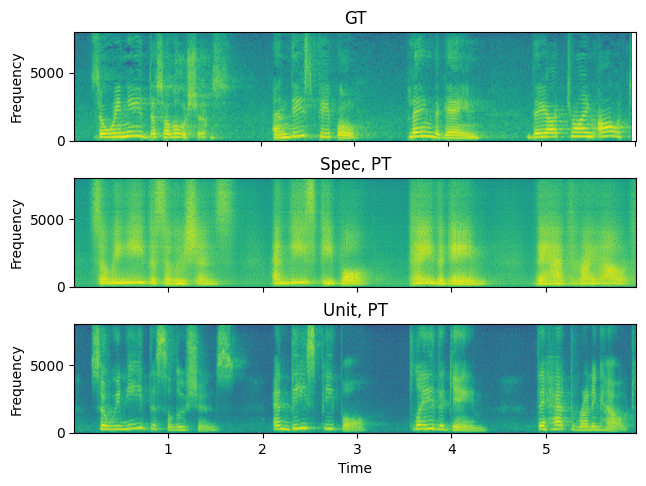}
    \caption{Spectrogram of ground truth (GT) and predicted audio generated by an ablated model that predicts spectrogram (Spec, PT) and the proposed ReVISE model (Unit, PT). We can observe that ``Spec, PT'' is more blurry, where harmonics (horizontal stripes) are missing.}
    \label{fig:spec-comparison}
\end{figure}

\cref{tab:abl_spec_and_pt} studies how these two factors together contribute to the superior performance observed from ReVISE. ``No-PT'' and ``PT'' indicates where pre-trained weights of AV-HuBERT is loaded, and ``Unit'' and ``Spec'' indicates the prediction target of the P-AVSR module. We also train a vocoder using LJSpeech to convert spectrogram to waveform for models that predict spectrograms. In terms of intelligibility, both pre-training and predicting units improve the performance, and pre-training is particularly important, similar to what prior studies observed on speech recognition~\cite{shi2022learning}.
Figure~\ref{fig:spec-comparison} shows the spectrograms of the three generated audios. Directly using spectrogram (Spec,PT) as prediction target leads to generation of more blurry speech compared to using units (Unit, PT).

\begin{table}[ht]
  \centering
  \begin{tabular}{@{}l|cc@{}}
    \toprule
    & \multicolumn{2}{c}{WER} \\
    Target & No-PT & PT \\
    \midrule\midrule
    Unit & 78.6\% & 35.5\% \\
    Spec & 96.9\% & 39.7\% \\
    \bottomrule    
  \end{tabular}
  \caption{Ablation studies comparing prediction target (unit vs. spec) and effectiveness of pre-training on different targets.}
  \label{tab:abl_spec_and_pt}
\end{table}

\subsection{On predicting low-level details}

\cref{tab:lrs3_v2a_app,tab:lrs3_inp_app,tab:lrs3_den_app,tab:lrs3_sep_app} show the complete results on the four LRS3 tasks, including metrics on intelligibility (WER), synchronization (LSE-C/D), quality (MOS), and low-level detail reconstruction (ESTOI and MCD). We see that while ReVISE outperforms or is on par with baseline methods on intelligibility, quality, and synchronization, it performs significantly worse than SVTS, Demucs, and VisualVoice on recovering low-level details. This is because ReVISE does not focus on reconstructing the exact reference signal; moreover, ReVISE does not aim to infer the speaker identity either since the SSL units are shown to encode very little speaker information.

In fact, it is not a deficiency that the proposed ReVISE model does not reconstruct reference signal exactly. As shown in \cref{tab:easycom-mos}, the quality of the reference signal may be mediocre for real world datasets such as EasyCom. In that scenario, reconstructing exact reference signal would lead to inferior performance compared to the proposed approach, where the training data for P-AVSR and P-TTS are decoupled, such that higher quality data can be used for the unit-to-speech module.

\begin{table}
  \centering
  \resizebox{\linewidth}{!}{
  \begin{tabular}{@{}l|r|rr|r|rr@{}}
    \toprule
    & Cont & \multicolumn{2}{c|}{Sync} & Qual & \multicolumn{2}{c}{Low-Level} \\
    Method & WER$\downarrow$ & LSE-C$\uparrow$ & LSE-D$\downarrow$ & MOS$\uparrow$ & ESTOI $\uparrow$ & MCD $\downarrow$ \\
    \midrule\midrule
    Tgt. audio & 5.6\% & 5.20 & 6.33 & 4.38\pmr{0.02} & - & - \\
    Tgt. audio resynthesis & 10.6\%	& 5.02 & 6.40 & 4.16\pmr{0.02} & 41.59 & 10.01 \\
    \midrule
    \multicolumn{7}{@{}l}{\textit{Silent video to speech}} \\
    \rowcolor{gray!10} SVTS-L, LRS3 & 81.5\% & 5.07 & 6.46 & 2.12\pmr{0.04} & 26.87 & 8.13 \\
    \rowcolor{gray!10} SVTS-L, LRS3+VoxEn & 67.4\% & 5.53 & 6.00 & 2.23\pmr{0.03} & 30.98 & 7.59 \\
    % ReVISE-Spec & V & \\
    ReVISE (Ours) & 33.9\% & 5.03 & 6.44 & 4.13\pmr{0.02} & 28.51 & 10.73 \\
    \bottomrule    
  \end{tabular}
  }
  \caption{Full video-to-speech synthesis results on LRS3 test set.}
  \label{tab:lrs3_v2a_app}
\end{table}

\begin{table}
  \centering
  \resizebox{\linewidth}{!}{
  \begin{tabular}{@{}lrr|r|rr|rr@{}}
    \toprule
    & & & Cont & \multicolumn{2}{c|}{Sync} & \multicolumn{2}{c}{Low-Level} \\
    Method & Inp Mod & Test & WER$\downarrow$ & LSE-C$\uparrow$ & LSE-D$\downarrow$ & ESTOI $\uparrow$ & MCD $\downarrow$ \\
    \midrule\midrule

    \rowcolor{gray!0} && 30\% & 44.1\% & 4.37 & 7.05 & 60.75 & 36.54 \\
    \rowcolor{gray!0} && 40\% & 53.5\% & 4.01 & 7.36 & 51.69 & 44.71 \\
    \rowcolor{gray!0} \multirow{-3}{*}{Inp. audio} & \multirow{-3}{*}{A} 
                       & 50\% & 60.3\% & 3.78 & 7.60 & 43.16 & 52.82 \\
    \midrule
                       
    \multicolumn{7}{@{}l}{\textit{Audio-visual speech inpainting}} \\
    
    \rowcolor{gray!10} && 30\% & 43.9\% & 4.37 & 7.04 & 59.40 & 35.95 \\
    \rowcolor{gray!10} && 40\% & 53.5\% & 4.04 & 7.34 & 50.55 & 43.96 \\
    \rowcolor{gray!10} \multirow{-3}{*}{Demucs} & \multirow{-3}{*}{A} 
                        & 50\% & 60.0\% & 3.82 & 7.57 & 42.20 & 52.09 \\

    \rowcolor{gray!0} && 30\% & 51.2\% & 4.20 & 7.14 & 25.27 & 11.57 \\
    \rowcolor{gray!0} && 40\% & 60.1\% & 3.94 & 7.38 & 21.55 & 11.91 \\
    \rowcolor{gray!0} \multirow{-3}{*}{Resynthesis} & \multirow{-3}{*}{A} 
                       & 50\% & 66.3\% & 3.68 & 7.63 & 17.97 & 12.09 \\

    \rowcolor{gray!10} && 30\% & 14.9\% & 5.01 & 6.42 & 38.15 & 10.23 \\
    \rowcolor{gray!10} && 40\% & 18.1\% & 4.99 & 6.44 & 37.05 & 10.31 \\
    \rowcolor{gray!10} \multirow{-3}{*}{ReVISE (Ours)} & \multirow{-3}{*}{AV} 
                        & 50\% & 19.2\% & 5.04 & 6.41 & 35.99 & 10.33 \\
      
    \bottomrule
    
  \end{tabular}
  }
  \caption{Full speech inpainting results on LRS3 test set with 30\%/40\%/50\% of frames dropped.}
  \label{tab:lrs3_inp_app}
\end{table}

\begin{table}
  \centering
  \resizebox{\linewidth}{!}{
  \begin{tabular}{@{}lrr|r|rr|r|rr@{}}
    \toprule
    & & & Cont & \multicolumn{2}{c|}{Sync} & Qual & \multicolumn{2}{c}{Low-Level}  \\
    Method & Mod & Test & WER$\downarrow$ & LSE-C$\uparrow$ & LSE-D$\downarrow$ & MOS$\uparrow$ & ESTOI $\uparrow$ & MCD $\downarrow$ \\
    \midrule
    \rowcolor{gray!0}Tgt. audio & - & - & 5.6\% & 5.20 & 6.33 & 4.36\pmr{0.018} & - & - \\
    \midrule\midrule
    \multicolumn{5}{@{}l}{\textit{Audio-visual speech denoising}} \\
    \rowcolor{gray!10} && lvl 1 &  7.8\% & 4.66 & 6.77 & 2.95\pmr{0.03} & 87.21 & 5.07 \\
    \rowcolor{gray!10} && lvl 2 & 20.0\% & 3.80 & 7.52 & 2.82\pmr{0.03} & 69.93 & 8.27  \\
    \rowcolor{gray!10} && lvl 3 & 63.9\% & 2.41 & 8.68 & 2.79\pmr{0.02} & 45.91 & 11.27 \\
    \rowcolor{gray!10} \multirow{-4}{*}{Inp. audio} & \multirow{-4}{*}{A}
                       & lvl 4 & 87.7\% & 1.54 & 9.39 & 2.71\pmr{0.04} & 25.55 & 13.47 \\
                       
    \rowcolor{gray!0} && lvl 1 & 13.3\% & 4.98 & 6.46 & 4.33\pmr{0.02} & 41.18 & 10.02 \\
    \rowcolor{gray!0} && lvl 2 & 25.9\% & 4.78 & 6.64 & 4.37\pmr{0.02} & 39.16 & 10.19 \\
    \rowcolor{gray!0} && lvl 3 & 69.4\% & 3.64 & 7.61 & 3.63\pmr{0.02} & 25.71 & 11.27 \\
    \rowcolor{gray!0} \multirow{-4}{*}{Resynthesis} & \multirow{-4}{*}{A}
                       & lvl 4 & 90.8\% & 2.11 & 8.85 & 3.34\pmr{0.03} & 9.20 & 12.56 \\
    
    \rowcolor{gray!10} && lvl 1 & 6.9\%  & 5.15 & 6.38 &3.58\pmr{0.04} & 92.59 & 3.73 \\
    \rowcolor{gray!10} && lvl 2 & 15.1\% & 4.95 & 6.57 &  3.63\pmr{0.03} & 85.61 & 4.81  \\
    \rowcolor{gray!10} && lvl 3 & 48.0\% & 4.14 & 7.28 & 3.36\pmr{0.03} & 65.94 & 6.81 \\
    \rowcolor{gray!10} \multirow{-4}{*}{Demucs~\cite{defossez2020real}} & \multirow{-4}{*}{A} 
                        & lvl 4 & 81.3\% & 2.43 & 8.75 & 2.89\pmr{0.04} & 31.52 & 10.25  \\
    
    \rowcolor{gray!0} && lvl 1  & 6.6\% & 5.16 & 6.35 & 4.11\pmr{0.03} & 92.20 & 3.89 \\
    \rowcolor{gray!0} && lvl 2 & 8.8\% & 5.19 & 6.35  & 3.94\pmr{0.02} & 86.92 & 5.07 \\
    \rowcolor{gray!0} && lvl 3 & 23.4\% & 5.07 & 6.45 & 3.80\pmr{0.02} & 74.16 & 6.45 \\
    \rowcolor{gray!0} \multirow{-4}{*}{VisualVoice~\cite{gao2021visualvoice}} & \multirow{-4}{*}{AV}
                        & lvl 4 & 58.0\% & 4.53 & 6.86 & 3.25\pmr{0.03} & 50.48 & 8.01 \\
    
    \rowcolor{gray!10} && lvl 1 & 9.4\%  & 5.03 & 6.42 & 4.31\pmr{0.02} & 41.22 & 10.04 \\
    \rowcolor{gray!10} && lvl 2 & 9.7\%  & 5.04 & 6.41 & 4.36\pmr{0.02} & 40.93 & 10.06 \\
    \rowcolor{gray!10} && lvl 3 & 11.7\% & 5.04 & 6.41 & 4.41\pmr{0.02} & 39.79 & 10.14 \\
    \rowcolor{gray!10} \multirow{-4}{*}{ReVISE (Ours)} & \multirow{-4}{*}{AV} 
                       & lvl 4 & 20.5\% & 5.07 & 6.40 & 4.34\pmr{0.02} & 35.80 & 10.36 \\
                       
    \bottomrule
  \end{tabular}
  }
  \caption{Full speech denoising results on LRS3 test set. SNR ranges for lvl 1/2/3/4 are [10,20]/[0,10]/[-10,0]/[-20,-10] dB.}
  \label{tab:lrs3_den_app}
\end{table}

\begin{table}
  \centering
  \resizebox{0.9\linewidth}{!}{
  \begin{tabular}{@{}lrr|r|rr|rr@{}}
    \toprule
    & & & Cont & \multicolumn{2}{c}{Sync} & \multicolumn{2}{c}{Low-Level} \\
    Method & Mod & Test & WER$\downarrow$ & LSE-C$\uparrow$ & LSE-D$\downarrow$ & ESTOI $\uparrow$ & MCD $\downarrow$ \\
    \midrule\midrule
    \multicolumn{5}{@{}l}{\textit{Audio-visual speech separation}} \\
    
    \rowcolor{gray!10} && lvl 1 & 15.7\% & 4.73 & 6.74 & 85.56 & 3.29 \\
    \rowcolor{gray!10} && lvl 2 & 65.7\% & 3.85 & 7.48 & 65.52 & 5.95 \\
    \rowcolor{gray!10} && lvl 3 & 102.3\% & 2.93 & 8.27 & 40.97 & 8.67 \\
    \rowcolor{gray!10} \multirow{-4}{*}{Inp. audio} & \multirow{-4}{*}{A}
                        & lvl 4 & 105.5\% & 2.46 & 8.66 & 22.33 & 10.51 \\
    
    \rowcolor{gray!0} && lvl 1 & 11.7  & 5.06 & 6.46 & 89.68 & 3.76 \\
    \rowcolor{gray!0} && lvl 2 & 55.7  & 4.10 & 7.25 & 66.80 & 6.27 \\
    \rowcolor{gray!0} && lvl 3 & 101.2 & 2.77 & 8.42 & 26.32 & 10.22 \\
    \rowcolor{gray!0} \multirow{-4}{*}{Demucs~\cite{defossez2020real}} & \multirow{-4}{*}{A}
                        & lvl 4 & 106.2 & 2.15 & 8.95 & 10.15 & 12.14 \\
    
    \rowcolor{gray!10} && lvl 1 &  6.2\% & 5.13 & 6.37 & 94.69 & 2.43 \\
    \rowcolor{gray!10} && lvl 2 &  7.8\% & 5.12 & 6.39 & 90.29 & 3.36 \\
    \rowcolor{gray!10} && lvl 3 & 15.3\% & 4.91 & 6.57 & 78.82 & 4.80 \\
    \rowcolor{gray!10} \multirow{-4}{*}{VisualVoice~\cite{gao2021visualvoice}} & \multirow{-4}{*}{AV}
                       & lvl 4 & 51.8\% & 4.20 & 7.17 & 53.84 & 7.02 \\
    
    \rowcolor{gray!0} && lvl 1 &  9.9\% & 5.05 & 6.41 & 41.34 & 10.05 \\
    \rowcolor{gray!0} && lvl 2 & 10.2\% & 5.05 & 6.41 & 40.90 & 10.07 \\
    \rowcolor{gray!0} && lvl 3 & 11.4\% & 5.05 & 6.41 & 40.07 & 10.11 \\
    \rowcolor{gray!0} \multirow{-4}{*}{ReVISE (Ours)} & \multirow{-4}{*}{AV} 
                        & lvl 4 & 15.7\% & 5.04 & 6.41 & 37.97 & 10.21 \\
    \bottomrule
    
  \end{tabular}
  }
  \caption{Full source separation results on LRS3 test set. SNR ranges for lvl 1/2/3/4 are [10,20]/[0,10]/[-10,0]/[-20,-10] dB.}
  \label{tab:lrs3_sep_app}
\end{table}

\section{Model Configurations}
\cref{tab:lrs3-param} and \cref{tab:easycom-param} detail the hyperparameters used for each experiment trained on LRS3 and EasyCom. Best checkpoints are selected based on the unit prediction accuracy on the validation set. Tri-stage learning rate schedules are used for all experiments, where the learning rate first ramps up linearly from 0 to the peak learning rate for the $t_1$\% of the total updates, remains at the peak learning rate for the next $t_2$\% of the total updates, and linearly decay to 5\% of the peak learning rate during the rest of the updates. We use $(t_1, t_2, 1 - t_1 - t_2)$ to denote the learning rate schedule.
Following \cite{baevski2020wav2vec}, we freeze the pre-trained module for a number of updates (num. of frozen steps) and only fine-tune the new modules (upsampling and softmax layers) at the beginning of training. We also follow~\cite{baevski2020wav2vec} to use SpecAug~\cite{park2019specaugment} for data augmentation and regularization, where random audio spans are dropped. The length of the spans and the ratio of frames dropped are labeled as ``audio masking length'' and ``audio masking prob'', respectively.

\begin{table}[htb]
  \centering
  \begin{tabular}{@{}l|c@{}}
    \toprule
    num. of updates & 15000\\
    num. of frozen steps & 0 \\
    tri-stage LR schedule & (33\%, 0\%, 67\%)\\
    peak learning rate   & 5e-5 \\
    audio masking prob & 0 \\
    audio masking length & n/a \\
    batch size / GPU & 1000 \\
    num. of GPU & 8 \\
    Adam $(\beta_1, \beta_2)$ & (0.9, 0.98) \\
    \bottomrule
  \end{tabular}
  \caption{EasyCom experiment hyperparameters.}
  \label{tab:easycom-param}
\end{table}

\begin{table*}[htb]
  \centering
  % \resizebox{\linewidth}{!}{
  \begin{tabular}{@{}l|ccccc@{}}
    \toprule
    & universal & video-to-speech & inpainting & speech denoising & source separation \\
    \midrule\midrule
    num. of updates & 45000 & 45000 & 45000 & 45000 & 45000 \\
    num. of frozen steps & 10000 & 5000 & 30000 & 5000 & 5000 \\
    tri-stage LR schedule & (10\%, 0\%, 90\%) & (10\%, 20\%, 70\%) & (10\%, 0\%, 90\%) & (10\%, 0\%, 90\%) & (10\%, 0\%, 90\%) \\
    peak learning rate   & 1e-4 & 6e-5 & 1e-4 & 1e-4 & 1e-4 \\
    audio masking prob & 35\% & n/a & 35\% & 35\% & 30\% \\
    audio masking length & 1 & n/a & 1 & 1 & 1 \\
    batch size / GPU & 1000 & 1000 & 1000 & 1000 & 1000 \\
    num. of GPU & 8 & 8 & 8 & 8 & 8 \\
    Adam $(\beta_1, \beta_2)$ & (0.9, 0.98) & (0.9, 0.98) & (0.9, 0.98) & (0.9, 0.98) & (0.9, 0.98) \\
    
    \bottomrule
    
  \end{tabular}
  % }
  \caption{LRS3 experiment hyperparameters.}
  \label{tab:lrs3-param}
\end{table*}

\end{document}